\documentclass[letterpaper, 10 pt, conference]{ieeeconf}  

\IEEEoverridecommandlockouts                              

\usepackage{graphicx} 
\usepackage{amsmath} 
\usepackage{amssymb}  

\title{\LARGE \bf
{\scshape i-GRIP}, a Grasping Movement Intention Estimator 
for Intuitive Control of Assistive Devices
}

\author{Etienne Moullet$^{1}$, Justin Carpentier$^{1}$, Christine Azevedo-Coste$^{2}$ and François Bailly$^{2}$
\thanks{$^{1}$E. Moullet and J. Carpentier are with WILLOW, INRIA Paris, Département Informatique
de l’ENS, Paris, France (corresponding author, e-mail: {\tt\small etienne.moullet@inria.fr}).}%
\thanks{$^{2}$C. Azevedo-Coste and F. Bailly are with CAMIN, INRIA centre d'Université Côte d'Azur, Université de Montpellier, Montpellier, France.}%
}

\begin{document}

\maketitle
\thispagestyle{empty}
\pagestyle{empty}

\begin{abstract}

This study describes and evaluates i-GRIP, a novel movement intention estimator designed to facilitate the control of assistive devices for grasping tasks in individuals with upper limb impairments. Operating within a collaborative grasping control paradigm, the users naturally move their hand towards an object they wish to grasp and i-GRIP identifies the target of the movement and selects an appropriate grip for the assistive device to perform. In an experimental study involving 11 healthy participants, i-GRIP exhibited promising estimation performances and responsiveness. The proposed approach paves the way towards more intuitive control of grasping assistive device for individuals with upper limb impairments. 

\end{abstract}

\section{INTRODUCTION}
    Various conditions - such as spinal cord injuries, stroke, or amputation - may hinder upper limb functions. Among them, grasping is crucial for many daily activities, and its impairment significantly impacts an individual's quality of life and autonomy. Various approaches and assistive devices aim to restore this function, including functional electrical stimulation (FES), exoskeletons, and prostheses. However, despite the continuous improvements in terms of actuation speed, accuracy, or strength, major challenges regarding user interfaces and control modalities remain \cite{gantenbein}.
    
    Regardless of the nature of the actuators, movement assistance is generally actively controlled by the user via dedicated sensors capturing the orders: joystick/switch buttons manipulation, neural activity (e.g., for Brain Computer Interface (BCI)), muscular activation (e.g., for myoelectric control), movements (e.g., using inertial measurements units (IMU)), voice, etc. \cite{azevedo, jiang_farina}. The choice of a given solution, driven by the residual functional capabilities of the user, the task to assist, and the device itself, has a major influence on the design of the control scheme of the device \cite{gantenbein}. 
    In the endeavor of restoring such a dexterous hand function as grasping, the design process of assistive solutions faces two challenges that are particularly hard to reconcile. On the one hand, the characteristics of objects, such as their shape, weight, and texture, can vary widely, requiring a high degree of flexibility and adaptability from the assistive device. On the other hand, individuals with upper limb disabilities often have a significantly reduced control over their body and thus struggle to control these devices efficiently~\cite{jiang_farina}. 
    
    Current control approaches often rely on state machine frameworks to accommodate for these conflicting constraints: the same user control input can trigger different device actions depending on the ongoing task.
    For example, in the case of trans-humeral amputation, the control of a prosthetic hand may be achieved by myoelectric control. Surface electrodes are placed on the upper arm muscles (the biceps and triceps) to detect their activation. The number of possible muscle contraction combinations being far inferior to the number of degrees of freedom (DOFs) of the prosthesis, the user needs to navigate between predefined modes in which a given muscular activation combination triggers, for instance, either the opening of the fingers or the extension of the wrist \cite{light}. Consequently, besides learning a dictionary of associations between muscle activation combinations and prosthesis action, a user needs to constantly switch between modes to achieve daily tasks.
    Attempts to circumvent these limitations focus on exploiting compensatory movements to control prostheses \cite{legrand}. Compensatory movements are intuitively elicited by the user to perform the motor task and interpreted as an input to control the prosthesis. While promising for single joint command (e.g., elbow, wrist), this approach seems difficult to adapt to adaptive tasks such as grasping.
   
    In the context of quadriplegia, hand function may be restored through FES: current pulses are applied via electrodes placed at the skin surface or in an implanted manner to activate peripheral motor nerves and evoke muscle contractions. In this context, control strategies may rely on switch buttons \cite{trotobas}, myoelectric control, inertial measurement units (IMU) installed on the contralateral shoulder, or even on voice control \cite{tigra}. Similarly to prosthesis control, such interfaces are associated to an important cognitive load on the user, in addition to lack of fluidity, resulting in sequential motions due to the state machine approach.   
    Stereo-vision and EMG have been used to drive a hand prosthesis in a semi-autonomous mode \cite{markovic_farina}, but required the user to select the target of his upcoming movement through visual feedback and muscle activation, hence disrupting the control flow.
    Alternative approaches, such as BCI, aim at directly detecting movement intent without relying on the interpretation of musculoskeletal actions \cite{nicolas_alonso}. Nevertheless, still to this day, these technologies often rely on sequential command paradigms or state machines \cite{padfield} and thus present the same limitations as previously stated. 
    In summary, while upper limb assistive technologies intrinsic performances may be considered satisfactory, existing command modalities often require the user to perform specific, sequential actions to control their device. Alas, these stereotypical motions result at best in an increased cognitive load exerted on the user, and at worst in saccadic movements that may hinder the realisation of their tasks in daily life. 
    
    In this paper, we present i-GRIP, a novel collaborative and generic grasping movement intention estimator, which adapts to the user's behavior without requiring any dedicated or stereotypical action. 
    The contributions of this paper follow its structure. First, we present the i-GRIP algorithm and its workflow. Then, we describe a practical implementation of i-GRIP, coupled with computer vision tools, and an experiment we ran to characterize its performances. Finally, we discuss our results and their implications.

    \begin{figure}
        \vspace*{2mm}
        \centering
        \includegraphics[width=\linewidth]{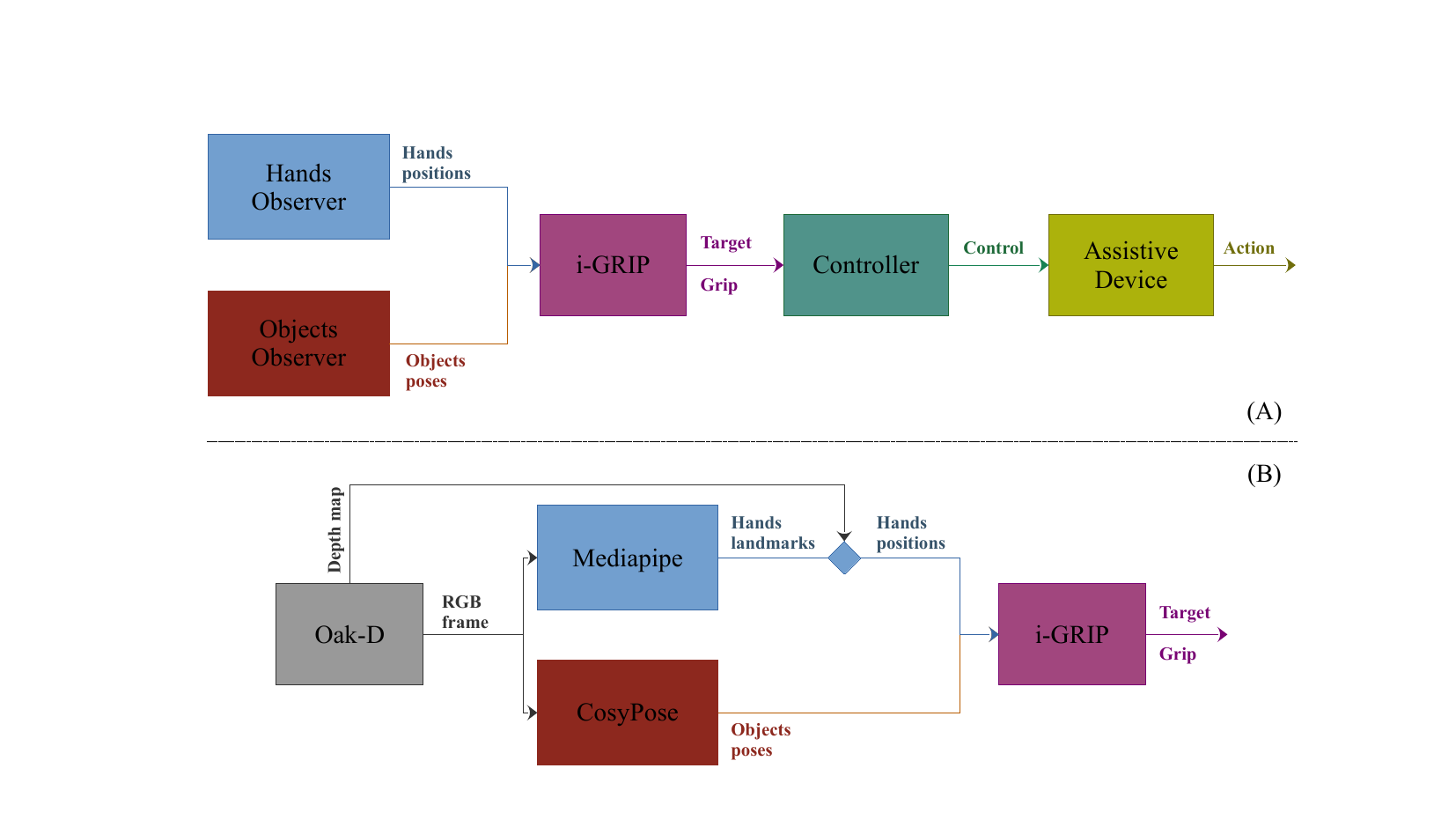}
        \caption{{\scshape i-GRIP} integration in a grasping assistive solution: (A) - in the generic case; (B) - in the demonstrator built for our experiment (see \ref{section:expe_choices})}
        \label{fig:i-grip_diagram_overview}
    \end{figure}

\section{{\scshape i-GRIP} algorithm}\label{section:i_grip}
    As previously exposed, a wide range of deficiencies may affect the upper limbs. A corresponding set of devices and sensors have been developed to assist with or restore grasping function. This work aimed to devise a real-time collaborative generic grasping movement intention estimator. 
    
    \subsection{General design principles}
    {\scshape i-GRIP} was designed to sit between the sensory and actuation functions to integrate it into various assistive solutions.      
    From continuous observations of a \textit{scene} (a human upper body and surrounding objects) i-GRIP gathers information about the initiation of a grasping movement by the user and transmits it to a dedicated, downstream process in charge of controlling an assistive device which terminates it autonomously (see Fig.\ref{fig:i-grip_diagram_overview}(A)). From these very general workflow principles, design choices were made regarding hands observation, objects observation, i-GRIP's outputs and general objectives.
    
    First, while real-time upper limb observations may rely on various sensing techniques, providing different levels of information \cite{bailly, yahya}, hand positions were chosen as the upper limb observations input of our algorithm. Indeed, they may be derived from the measurements of diverse kind of sensors and are thus compatible with different user impairments. Moreover, such low complexity measurements were deemed less susceptible to be perturbed by the actions of the assistive device (as full hand joint poses would be, for instance).
    Next, surrounding objects observations inputs were chosen to be their 6d poses in order to properly leverage their shapes.    
    Then, in order to accommodate for diverse assistive devices, i-GRIP output was meant to be high level information about the initiation of a grasping movement by the user: the target of its movement and an appropriate grip to grasp it.   
    Finally, i-GRIP algorithm was conceived to interpret natural movements from the user, requiring little to no adaptation.
    For this purpose, the grip selection process was developed to be easily understandable by the user and facilitate the anticipation of the device's actions (see \ref{section:grip_selection}).     
    
    \subsection{Internal workflow overview}
\begin{figure*}
        \centering
        \vspace*{2.mm}
        \includegraphics[scale =0.90]{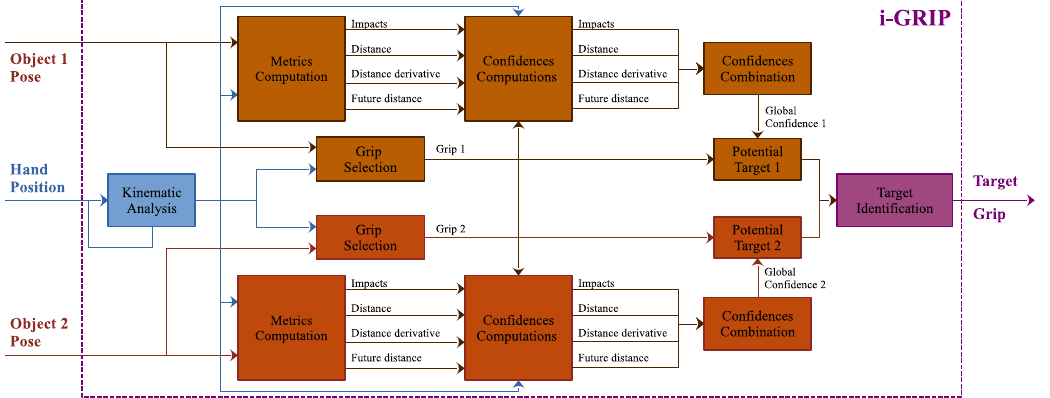}
        \caption{Data workflow and computations inside i-GRIP for one hand and two objects.}
        \label{fig:i-grip_diagram_complete}
    \end{figure*}

    {\scshape i-GRIP} is able to track the movements of several hands at the same time, but their analysis is entirely parallelized and no bimanual task is taken into account. We thus present in the following and in Fig. \ref{fig:i-grip_diagram_complete} the processing steps for a single hand.
    Whenever a new hand or objects observation is sent to i-GRIP, a full loop of computation is ran. It first performs kinematic analyzes of the hands motions to predict their near future trajectories. Next, every detected object is treated as a potential target and a confidence score is computed for each hand-object pair. Then, the object carrying the highest confidence score is identified as the target of the movement. Finally, the position of the hand relative to the object and its shape allows for the selection of an appropriate grip. Fig. \ref{fig:i-grip_diagram_complete} illustrates the data flow and computation steps performed by i-GRIP, which are further detailed in the following subsections.
    
    Please note that every formula presented in the rest of this section corresponds to a single loop of i-GRIP. The time variable was not included for the sake of notations clarity.
 
    \begin{figure}
        \centering
        \includegraphics[scale=0.1]{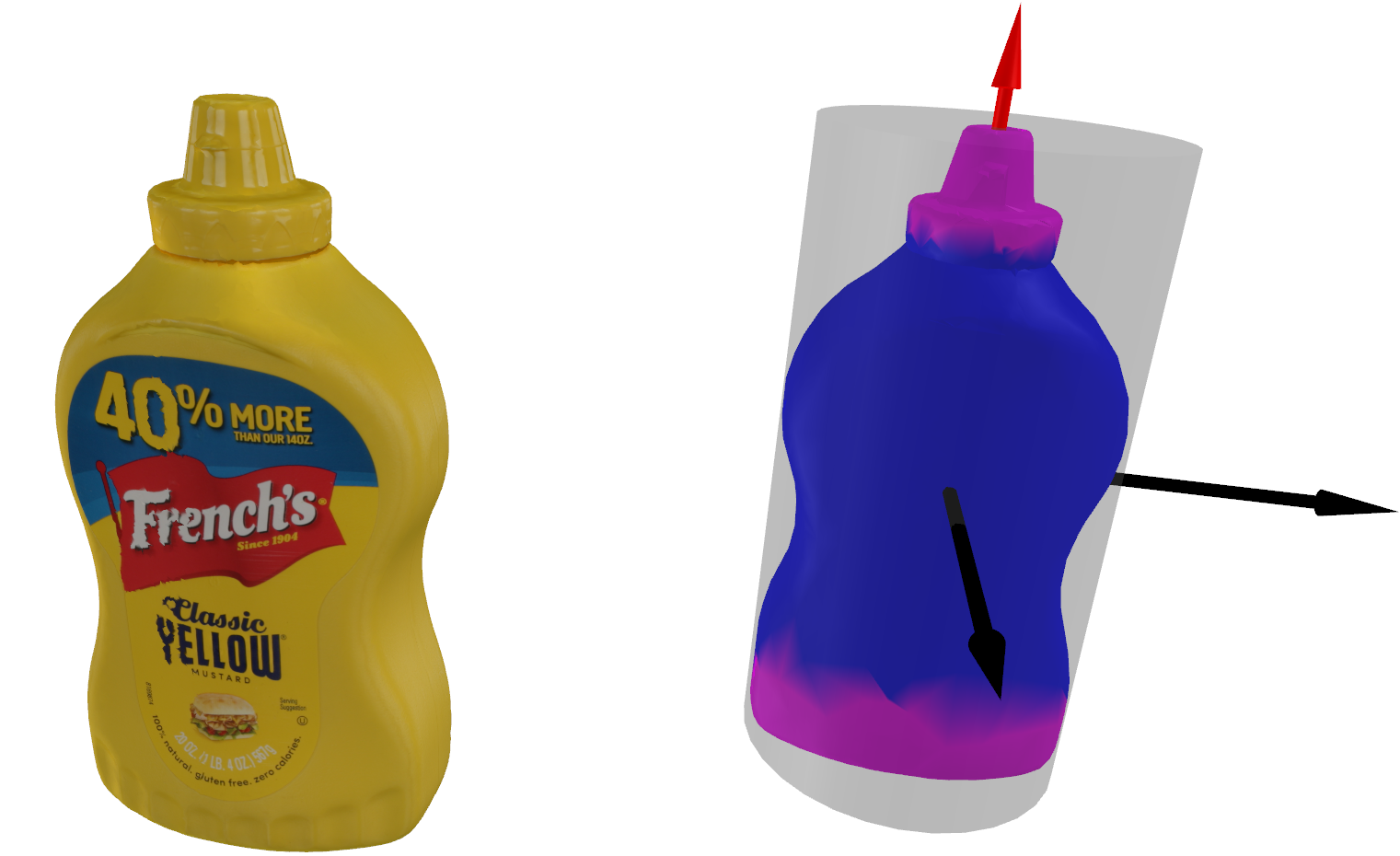}
        \caption{Mustard bottle from YCB set \cite{ycb} (left) and a visualization of the corresponding grip selection process (right). Transparent grey volume is the bounding cylinder of the mesh. The red arrow figures the z-axis of the object’s frame. The blue zone illustrates z-values corresponding to a palmar grip, while the magenta zone and outwards correspond to a pinch grip.}
        \label{fig:grip}
    \end{figure}
   
    \subsection{Scene observations pre-processing}\label{section:observations_pre_processing}
    As previously stated, i-GRIP is designed to work downstream of two scene observation processes that provide hands and objects observations as inputs. More precisely, those inputs consist in: 
    \begin{itemize}
        \item a list of hands detected in the observed scene $\boldsymbol{\mathcal{H}} =\left\{\boldsymbol{h}^i, i \in \mathbb  {N}  \right \}$, embedding their 3d positions $\boldsymbol{p}^i$ and measurement timestamps $t^i$,
        \item a list of objects detected in the observed scene $\boldsymbol{\mathcal{O}} =\left\{\boldsymbol{o}^j, j \in \mathbb  {N}  \right \}$, embedding their 6d poses (3d positions $\boldsymbol{p}^j$ and 3d orientations $\boldsymbol{\theta}^j$) as well as their 3d mesh models $\boldsymbol{M}^j$  and measurement timestamps $t^j$.
    \end{itemize}
    
    In a preliminary step, $\boldsymbol{p}^i$, $\boldsymbol{p}^j$ and $\boldsymbol{\theta}^j$ are filtered using speed-based low-pass filter, in order to mitigate the impact of the noise in the received measurements,.
    Then, a kinematic analysis of hands' movements is performed. We keep track of the recent past trajectory of a hand $\boldsymbol{h}^i$ in a sliding window $\boldsymbol{P}^i = \left [\left( \boldsymbol{p}_{-k}^i,\tau_k^i \right) \right ], k \in \left [ 0,n_f \right ]$, where $\tau_k = t^i - t_k^i$ is the time elapsed between the last available measurement and the measurement of $\boldsymbol{p}_{-k}^i$, and where $n_f$ is a fixed size adapted to the frequency of acquisition of inputs. We next perform a weighted second order polynomial regression on this window, with weights defined as 
    \begin{equation}
        w_k = \frac{1}{1+\tau_k}
    \end{equation}
    Then, we compute a list of future trajectory points $\boldsymbol{P}_{future}^i$ by extrapolating the computed polynomial to near-future timestamps. These variables are used to achieve target identification and grip selection.

    \subsection{Object metrics evaluation}

         A 3D virtual scene (see Fig. \ref{fig:setup}-B) incorporates the processed hand $\boldsymbol{h}^i$ as a 3D point $\boldsymbol{p}^i$ and each processed object $\boldsymbol{o}^j $ as its mesh $\boldsymbol{M}^j$ placed according to its pose $(\boldsymbol{p}^j, \boldsymbol{\theta}^j)$.
        For a given hand $\boldsymbol{h}^i \in \boldsymbol{\mathcal{H}}$, every object $\boldsymbol{o}^j \in \boldsymbol{\mathcal{O}}$ detected in the scene is treated as a potential target and processed separately. 
         For each pair $(\boldsymbol{h}^i | \boldsymbol{o}^j)$, four different metrics are computed:
        \begin{itemize}
            \item \textbf{ray impacts number}: from predicted future points $\boldsymbol{P}_{future}^i$ we cast cones of rays whose axes are co-linear to the local velocity vector, and whose angles depend on the local scalar velocity (see Fig. \ref{fig:setup}). The metric $n_{impacts}^{i,j}$ then consists in the number of rays that intercept the object's mesh.
            
            \item \textbf{distance derivative}: we define $d_{center}^{i,j} =  \left\|\boldsymbol{p}^i - \boldsymbol{c}^j \right\| $ as the distance between the hand's position and the position of the center of the object's mesh $ \boldsymbol{c}^j $. We define this metric as its time derivative $d_{der}^{i,j} = \frac{d}{dt}d_{center}^{i,j}$ 
            
            \item \textbf{distance}: we define this metric as the distance between the hand's position and the closest node on the object's mesh $d_{mesh}^{i,j} = \underset{\boldsymbol{n}_l^j \in \boldsymbol{M}^j}{\min} \left(  \left\|\boldsymbol{p}^i - \boldsymbol{n}_l^j \right\| \right)$.
            
            \item \textbf{future distance}: we denote $\boldsymbol{p}_{future,m}^i$ as the barycenter of the predicted future points $\boldsymbol{P}_{future}^i$. We define this metric as its distance to the object's mesh  $d_f^{i,j} = \underset{\boldsymbol{n}_l^j \in \boldsymbol{M}^j}{\min} \left(  \left\|\boldsymbol{p}_{f,m}^i - \boldsymbol{n}_l^j \right\| \right)$. 
            
        \end{itemize}

        \subsection{Object metrics confidence scores}
        At this step of the algorithm, we put the metrics defined above back in perspective of the whole scene: taking into account the hand's absolute motion and the other objects' positions, we compute four confidence scores that evaluate the likeliness of an object to be the target of the movement according to the corresponding metric.
        
        \begin{itemize}
            \item \textbf{ray impacts number}: we denote $n_{impacts}^{i,tot} = \sum_{j}^{o} {n_{impacts}^{i,j}}$ as the total amount of impacts over all detected objects and define the impacts confidence score as 
                \begin{equation}\label{eq_confidence_impacts}
                    c_{impacts}( \boldsymbol{h}^i| \boldsymbol{o}^j) =  \frac{n_{impacts}^{i,j}}{n_{impacts}^{i,tot}}
                \end{equation}
            \item \textbf{distance derivative}: we denote $v^i$ as the hand's scalar velocity and define the distance derivative confidence score as
                \begin{equation}\label{eq_confidence_distance_derivative}
                    c_{d\_der}( \boldsymbol{h}^i| \boldsymbol{o}^j)= \frac{d_{der}^{i,j}}{v^i}
                \end{equation}
            \item \textbf{distance}: we denote $d_{max}$ as the maximum distance between all pairs of potential targets, and define the distance confidence score as
                \begin{equation}\label{eq_confidence_distance}
                    c_{dist}( \boldsymbol{h}^i| \boldsymbol{o}^j) = 1 - \max \left ( \frac{d_{mesh}^{i,j}}{d_{max}}, 1 \right )
                \end{equation}
            
            \item \textbf{future distance}: similarly to \eqref{eq_confidence_distance}, we define the corresponding confidence score as
                \begin{equation}\label{eq_confidence_future_distance}
                    c_{fut\_dist}( \boldsymbol{h}^i| \boldsymbol{o}^j) = 1 - \max \left ( \frac{d_f^{i,j}}{d_{max}}, 1 \right )
                \end{equation}
            
        \end{itemize}

        \subsection{Object global confidence score}
        We assign to each metric $m \in M  = \left \{ impacts, d\_der, dist, fut\_dist \right \}$ a corresponding weight $w_m$. We define variable weights $w_0$ and $w_1$, denoting that they apply respectively to zero and first order derivative metrics, so that $w_{distance} = w_{future\_distance} = w_{0}$ and $w_{impacts} = w_{\text{distance\_derivative}} = w_{1}$. We use the fact that during reaching tasks human hand's scalar velocity consistently follows a well-known "bell-shaped curve" \cite{flash_hogan} (see Fig. \ref{fig:confidences}(A) for an illustration from our data). We detect whether the hand is before velocity peak or after by comparing the current scalar velocity to the recent-past maximum scalar velocity $v_{max\_win} = \max \left( V^i \right)$ where $V^i = \left [ v_{-k}^i \right ], k \in \left [ 0,n_f \right ]$ is a sliding window of fixed size $n_f$ of the hand's scalar velocity. We compute $w_0$ as 
        \begin{equation}
            w_0 = \frac{v^i}{v_{max\_win}}
        \end{equation} 
        and $w_1$ as
        \begin{equation}
        w_1 = \left \{
            \begin{array}{ccc}
                 2 & if & v^i > v_{max\_win}  \\
                 1 & if & v^i \le v_{max\_win}  \\            
            \end{array}
        \right .
        \end{equation}
        We then compute a global target confidence as the weighted sum of these metrics confidences:
        \begin{equation}\label{eq_global_confidence}
            c_{glob}( \boldsymbol{h}^i| \boldsymbol{o}^j) = \frac{\sum_{m \in M}w_{m}c_{m}( \boldsymbol{h}^i| \boldsymbol{o}^j) }{\sum_{m \in M}w_{m}} 
        \end{equation}
        
    \begin{figure}
        \centering
        \includegraphics[width = \linewidth]{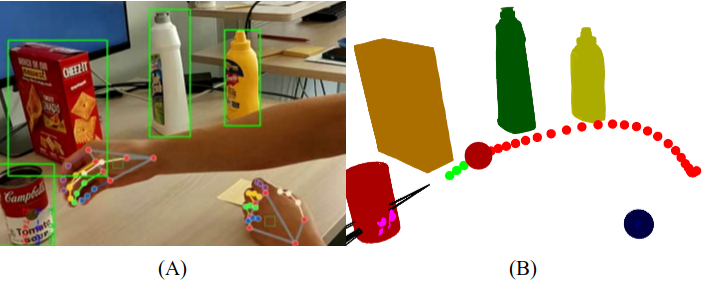}
        \caption{Recorded scene from the experiment and its analysis by i-GRIP: (A) - Example of a video frame captured during a trial overlaid with green rectangles marking the detected objects and multicolored landmarks marking hands key-points. (B) - Corresponding 3D virtual scene: Orange, yellow, red, and green objects are the rendered meshes of the detected objects. Big blue and red spheres represent the 3D positions of, respectively, left and right hands. The middle-sized red and green spheres represent, respectively, the past and expected future trajectory of the right hand. Black lines are a cones of rays expanding from the hands' expected future trajectories, whose impacts on the objects' meshes are the small magenta dots. }
        \label{fig:setup}
    \end{figure}
    
        \subsection{Target identification}
        When the previously described steps have been performed for every detected object $\boldsymbol{o}^j \in \boldsymbol{\mathcal{O}}$, we synthesize and compare their respective global confidences. The most probable target for the motion of hand $ \boldsymbol{h}^i$ is then defined as the potential target with highest global confidence:
    
        \begin{equation}\label{eq_target_identification}
            target(\boldsymbol{h}^i) = \underset{\boldsymbol{o}^j}{\arg \max}(c_{glob}( \boldsymbol{h}^i| \boldsymbol{o}^j))
        \end{equation}
        
        For the sake of comparison (see \ref{section:results}), we also define target as it would be found considering each metric separately for $m \in M $:  
        \begin{equation}\label{eq_target_identification_by_metric}
            target_{m}(\boldsymbol{h}^i) = \underset{\boldsymbol{o}^j}{\arg \max}(c_{m}( \boldsymbol{h}^i| \boldsymbol{o}^j))
        \end{equation}

    \subsection{Grip selection}\label{section:grip_selection}
        In its current version, i-GRIP is focused on oblong objects, that may be grasped using two of the main grips used in daily life \cite{feix}: palmar and pinch grips. For each object, we define a reference frame $R^j$ whose origin is the center of gravity of its mesh, and oriented such as the z-axis matches the axis of the bounding cylinder of the mesh. We denote $L^j$ the length of this cylinder.  During the movement of a hand $ \boldsymbol{h}^i$ towards an object $\boldsymbol{o}^j$, its position is expressed in the object's frame as $\boldsymbol{p}^{i,j}$, and its z-component is compared to $L^j$ to determine the appropriate grip as 
        \begin{equation}
        grip (\boldsymbol{h}^i| \boldsymbol{o}^j) =\left \{
            \begin{array}{ccc}
                  \text{pinch} & if & | z^{i,j} | \geq 0.85*L^j  \\
                  \text{palmar} & if & | z^{i,j} | < 0.85*L^j  
            \end{array}
        \right .
        \end{equation}
        Fig. \ref{fig:grip} gives a visualization of this selection process. The most probable appropriate grip for a given hand is then defined as the grip corresponding to the most probable target in \eqref{eq_target_identification}:
        \begin{equation} \label{eq_grip_selection}
             grip(\boldsymbol{h}^i)=grip(\boldsymbol{h}^i| target(\boldsymbol{h}^i))
        \end{equation}

       \begin{figure*}
       \centering
       \vspace*{2mm}
          \includegraphics[scale=0.92]{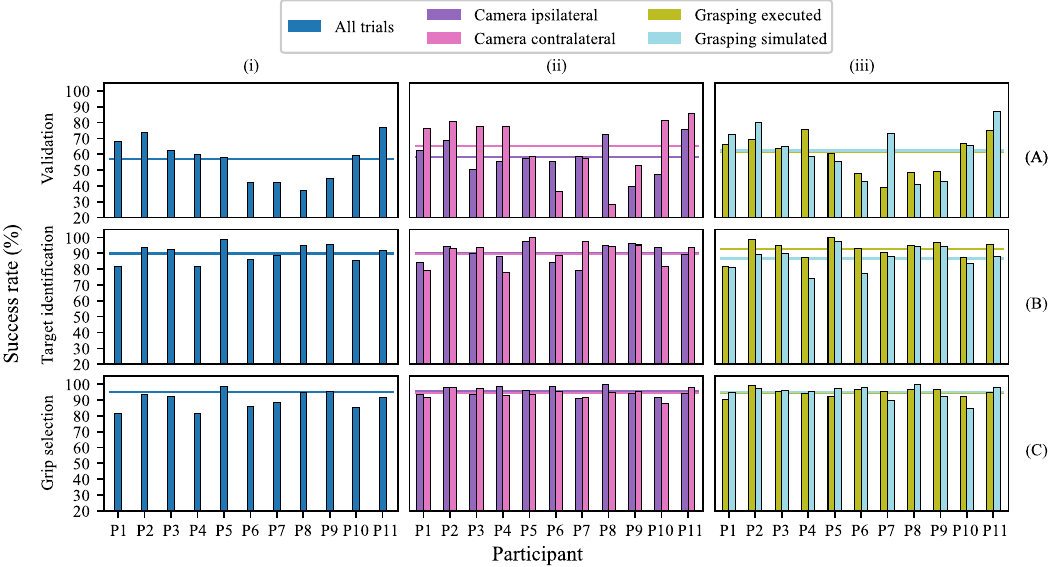}
          \caption{Processing success rates for each participant under different conditions. Rows (A), (B) and (C) show respectively video trials validation, target detection and grip identification success rates. Columns (i), (ii) and (iii) correspond respectively to all trials, trials regrouped according to relative camera placements and trials regrouped according to grasping movement types. Horizontal colored lines represent the average rates over all trials for a given condition.}
          \label{fig_eval_participants}
       \end{figure*}
       
\section{Experimental evaluation}
    As a reminder (see \ref{section:i_grip} and Fig. \ref{fig:i-grip_diagram_overview}(A)), i-GRIP is not a controller, but rather provides target and grip information to a downstream process in charge of controlling an assistive device. 
    Thus, the assessment of i-GRIP was restricted to its outputs and an experiment was conducted to gather a dataset of hand trajectories and their corresponding "ground truths", consisting in the targets the movements were directed to and the grips that were employed. 
    
    \subsection{Observation processes experimental implementation}\label{section:expe_choices}
        Even though no assistive device was included in the experiment, i-GRIP is not a standalone solution and expects hands positions and objects poses as inputs. An implementation of the observing processes presented in Fig. \ref{fig:i-grip_diagram_overview}(A) were thus required to build a demonstrator and evaluate i-GRIP. 
        In order to approach real-life conditions and even though data analysis would be done offline, only real-time solutions were considered for data acquisition.  
        Prioritizing cost, ease of use, installation and universality, stereoscopic RGB-d cameras (OAK-D S2, Luxonis) were chosen as the primary observation system. 
        
        In coherence with the chosen sensors, Mediapipe \cite{mediapipe} and CosyPose \cite{cosypose} were used to extract respectively the positions of hands and poses of objects present in the filmed scene. 
        Mediapipe \cite{mediapipe}, was used on RGB frames to get 2D hand landmarks. Depth maps were then leveraged to compute 3D coordinates of each landmark. Finally, hands 3D positions were measured at the center of the palm. 
        CosyPose \cite{cosypose} was fed RGB frames and allowed to detect objects and estimate their 6D poses. The version that was used was specially trained on the YCB set \cite{ycb}, from which objects were picked for the experiment (see \ref{section:expe_proc}). 
        Fig. \ref{fig:i-grip_diagram_overview}(B) illustrates the data workflow of the built demonstrator.
       
    \subsection{Experimental setup and procedure}\label{section:expe_proc}
        An experimental study approved by INRIA ethical committee (COERLE Decision 2024-01) involving eleven healthy participants was conducted to evaluate i-GRIP algorithm performances. 
        Participants were seated in front of a table. Four daily life objects were selected from the YCB set \cite{ycb}: a mustard bottle, a bleach bottle, a tomato can or a box of cheez'it. They were randomly placed between the participant's hands resting position and a computer screen on which were displayed instructions (see Fig. \ref{fig:setup}(A)). These instructions informed the participant on the four settings that characterized the upcoming trial he had to perform:
        \begin{itemize}
            \item \textit{hand}: whether to use the left or right hand,
            \item \textit{target}: which object to aim at among the 4 objects on the table
            \item \textit{grip}: whether to apply a pinch or palmar grip,
            \item \textit{movement type}: whether to execute the grip (actually grasping the object) or simulate it (approaching the object without moving their fingers during movement or grasping the object).
        \end{itemize}
        Four repetitions of each possible combination of these settings, leading to 128 trials, were randomly shuffled for each participant.
        Two RGB-d cameras (OAK-D S2, Luxonis) were placed on the left and right sides of the participant, at shoulder level, and filmed both of their hands and every object on the table at 30 frames per second (FPS). The two resulting points of view doubled the number of video trials to analyse.

        \subsection{Recordings pre-processing}\label{section:recordings_pre_processing}
         Recordings were cut to match the beginning and end of each trial, checked to exclude trials where participants didn't perform the required combination.
        
        Preliminary testing of the observation tools described in \ref{section:expe_choices} exhibited inconsistent performances on videos and that hands and/or objects detections were not successful on all frames. Being based on a kinematic properties, i-GRIP requires a minimal temporal continuity in the inputs it is fed to allow for a relevant analysis. 
        Trials pre-processings results were thus analysed to verify if detections were consistent enough throughout their videos. A first validation condition verified if hands and objects first detections happened early enough during the trial. A second validation condition verified if, once detected, hands and objects detections did not fail for too many consecutive frames.
        Trials that met both conditions were deemed valid, while the others were excluded from the following analysis.

        \subsection{Trial evaluation}
        The i-GRIP workflow (see \ref{section:i_grip}) was applied to each recorded RGB-d frame, providing the corresponding target and grip predictions sequences, that were compared to the ground truth of the trial instructions. 
        
        In order to evaluate target identifications throughout a whole valid video trial, $k_{target}$ denoted the index of the first frame of the longest sequence of frames for which target identification was successful, and $k_f$ denoted the index of the end of movement frame. $n_{target}$ denoted the amount of frames between $k_{target}$ and $k_f$ for which target identification was successful and the success ratio $r_{target}=\frac{n_{target}}{k_f - k_{target}}$ was computed. A trial was deemed successful in the regard of target identification if $r_{target} > 70\%$. 

        Similarly to target identification, $k_{grip}$ identified the index of the first frame of longest sequence of frames for which target identification was successful, and $k_f$, the index of the end of movement frame. $n_{grip}$ denoted the amount of frames between $k_{grip}$ and $k_f$ for which target identification was successful and the success ratio $r_{grip}=\frac{n_{grip}}{k_f - k_{grip}}$ was computed. A trial was deemed successful in the regard of grip selection if $r_{grip} > 70\%$. 
        
        Temporal delay and margin were computed respectively as the time elapsed between the movement onset and the target detection, and between the target detection and the end of the movement. Similar computation was performed for grip selection.

       \begin{figure}
           \centering
           \includegraphics[width = \linewidth]{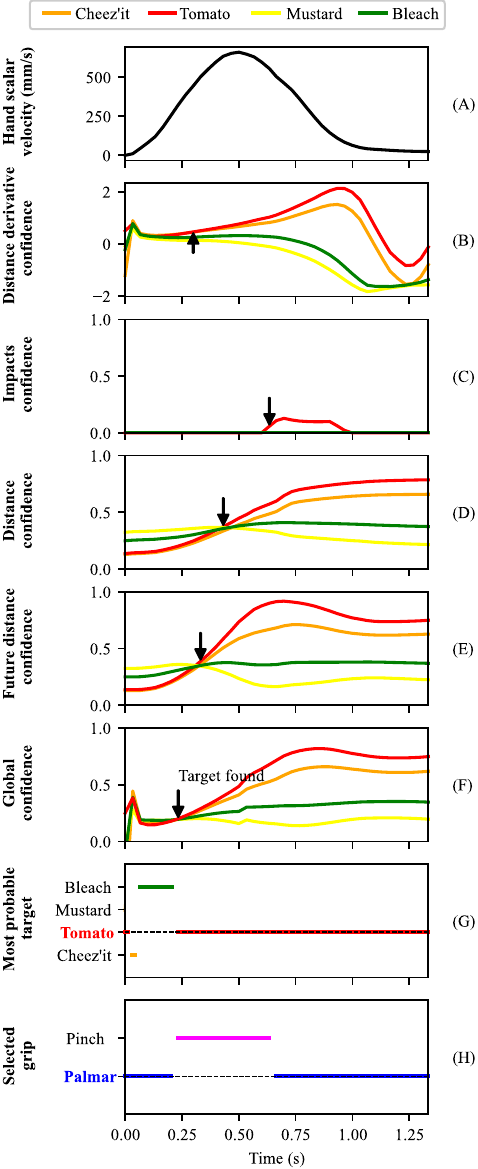}
           \caption{Temporal data of a typical video trial from participant P2, recorded from the camera in contralateral position. The task was to reach for the tomato can with their right hand and execute a palmar grip. (A) shows the hand's velocity profile. (B) to (F) show the temporal evolution of the target metrics confidences defined in \eqref{eq_confidence_impacts}, \eqref{eq_confidence_distance_derivative}, \eqref{eq_confidence_distance}, \eqref{eq_confidence_future_distance} and \eqref{eq_global_confidence}. Black arrows show the moments where each metric found the correct target. (G) and (H) illustrate the resulting identified target and selected grip over time.
           }
           \label{fig:confidences}
       \end{figure}

    \section{Results}\label{section:results}
    
    Hands and objects observation processes upstream of i-GRIP were applied on a total of 74378 frames and were fully successful on 66499 of them (89.4\%). 
    One participant could not pursue the experiment until the end, and the data of four trials were corrupted, resulting in a total of 1359 performed trials. Out of the corresponding 2718 recorded video trials, 1553 were deemed valid (Sec. \ref{section:recordings_pre_processing}). Row A in Fig. \ref{fig_eval_participants} shows the validation rates across participants for all trials under different conditions. All results displayed in the following subsections only relate to valid trials.
    Fig. \ref{fig:confidences} shows the temporal evolution of i-GRIP's metrics for a valid and representative trial.

    \subsection{Target identification}
        
    Target identification following the global confidence defined in \eqref{eq_target_identification} was successful on 61.9\% of the frames of valid video trials. Frame-by-frame success rates of individual metrics defined in \eqref{eq_confidence_impacts}, \eqref{eq_confidence_distance_derivative}, \eqref{eq_confidence_distance} and \eqref{eq_confidence_future_distance} are given in the first column of Table \ref{table_target_identification}. 

    Target was successfully identified on 89.9\% of all valid video trials. Trial success rates of the individual metrics are given in the second column of Table \ref{table_target_identification}. When computed for each participant over every recording devices, the average success rate was 90.1\% with a standard deviation of 7.5\% (see Fig. \ref{fig_eval_participants}(B) for the target identification success rate across participants and different experimental conditions). 
    Video trial target identification success rate reached: 
    \begin{itemize}
            \item 90.5\% with the camera placed on the ipsilateral side of the moving hand, and 89.4\%  with the one placed on the contralateral side, 
            \item 92.9\% when the grip was executed at the end of the movement, and 86.8\% when it was not.
    \end{itemize} 

    Targets were correctly identified within an average delay of 0.52s, resulting in an average temporal margin until the end of the movements of 0.67s. Table \ref{table_target_identification} gives the target detection delays and margins for every confidence in the third and fourth columns respectively.

    \subsection{Grip selection}
    Grip was successfully selected on 94.8\% of all video trials. When computed for each participant over every recording device, the average success rate was 94.5\% with a standard deviation of 4.3\% (see Fig. \ref{fig_eval_participants}(C) for the grip selection success rate across participants and different experimental conditions). Grip selection success rate reached: 
    \begin{itemize}
        \item 95.5\% with the camera placed on the ipsilateral side of the moving hand, and 94.2\%  with the one placed on the contralateral side, 
        \item 94.7\% when the grasp was executed at the end of the movement, and 94.9\% when it was not.
    \end{itemize} 
    Grips were correctly selected within an average delay of 0.39s, resulting in an average temporal margin until the end of the movements of 0.80s.

\section{Discussion}
In this paper, we presented i-GRIP, an algorithm conceived to estimate grasping movement intention. {\scshape i-GRIP} was designed to be integrated into a wide variety of assistive solutions and apparatuses (see Fig. \ref{fig:i-grip_diagram_overview}). Its expected inputs were limited to estimations of the positions of hands and poses of objects observed in a monitored scene. Its outputs, to be transmitted to a controlled upper limb assistive device, consisted in the identification of the object targeted by an ongoing movement and an appropriate grip to grasp it. As i-GRIP treats hands separately, we described the computational loop workflow for a single hand, and showed its illustration for two objects in Fig. \ref{fig:i-grip_diagram_complete}. First, a kinematic analysis of a hand's absolute trajectory was performed, to predict its near future motion. Then, every detected object was treated as a potential target, and an analysis of the hand's motion relative to each of them was conducted: four metrics and their corresponding confidence scores were computed in \eqref{eq_confidence_impacts}, \eqref{eq_confidence_distance_derivative}, \eqref{eq_confidence_distance} and \eqref{eq_confidence_future_distance}, and were then synthesized into a global confidence score in \eqref{eq_global_confidence}, evaluating the overall likeliness of an object being the target of the movement. In the meantime, the shape of each object was leveraged to select an appropriate grip according to the hand's position relative to them. Finally, the target output was identified as the object presenting the highest global confidence score in \eqref{eq_target_identification} and the grip output was set as the grip embedded by the target output in \eqref{eq_grip_selection}.

    \begin{table}[h]
    \caption{Metrics target identification accuracy (success rates for single frames and whole trials) and reactivity (delays and margins over trials)}
    \label{table_target_identification}
    \begin{center}
    \begin{tabular}{|c||c|c|c|c|}
    \hline
    Metric & Frames & Trials & Delay (s) & Margin (s)\\
    \hline
    \hline
    Global & 61.9\% & 89.9\% & 0.52& 0.67 \\
    \hline
    Distance derivative & 56.8\% & 72.0\%& 0.48 & 0.72\\
    \hline
    Impacts & 15.7\% & 34.7\%& 0.97 & 0.25\\
    \hline
    Distance & 53.6\% & 88.2\%& 0.59 & 0.60\\
    \hline
    Future distance & 59.6\% & 88.7\% & 0.52 & 0.58\\
    \hline
    \end{tabular}
    \end{center}
\end{table}

\subsection{Overall performances}
{\scshape i-GRIP}'s success rates for target identification and grip selection over trials, computed over all participants, trial settings and recording devices, reached respectively 89.9\% and 94.8\%. 
The success rates presented for each participant in Fig. \ref{fig_eval_participants}(B,i) and (C,i) and their reported corresponding mean values and standard deviations show inter-participant variability, suggesting that improvement could be achieved in hand-tailoring i-GRIP specifically for a user. 

Additionally, it may be noticed in Fig. \ref{fig_eval_participants} and the reported results that the grip selection success rate is greater than the target identification success rate for each experimental condition, which is also the case with all video trials combined. 
Similarly, reported average delays and margins computed over all participants, trial settings and recording devices, show that grip selection was quicker than target identification.
This means that while the target was misidentified, the grip associated to it was correct which may be explained by the similar oblong shape the objects shared. As a consequence, adaptations might be necessary when expanding i-GRIP to different shapes or grip types.

\subsection{{\scshape i-GRIP} workflow analysis}
Fig. \ref{fig:confidences} shows the evolution during a typical trial of the confidence scores in \eqref{eq_confidence_impacts}, \eqref{eq_confidence_distance_derivative}, \eqref{eq_confidence_distance}, \eqref{eq_confidence_future_distance} and \eqref{eq_global_confidence} for the four detected objects, as well as i-GRIP's corresponding outputs.  
One can notice that every metric confidence scores fails to identify the target at the beginning of the movement. In fact, Table \ref{table_target_identification} shows that they are only able to identify a target after some delay (see arrows in Fig. \ref{fig:confidences} for a visualization in a single trial). This can be explained by the fact that they all rely on the observation of the hand's trajectory, and cannot anticipate the user's intent before it is translated into a distinguishable motion. Yet, one can notice that the metrics complement each other: "velocity-related" metrics (i.e impacts and distance derivative) tend to react faster than "distance-related" metrics (i.e distance and future distance), but the latter takeover in terms of confidence toward the end of the movement. 

Moreover, interestingly, as shown in Table \ref{table_target_identification}, individual metrics under-performed the global confidence, both frame-by-frame and for video trials. This means that while an object might not bear the maximum metric confidence score, the combination of confidence scores into the global one allows i-GRIP to identify it as the correct target. Additionally, Table \ref{table_target_identification} illustrates that the global confidence achieves timings close to the best performing metric on this matter (i.e distance derivative), all while being superior in terms of success rates in target identification and grip selection.

\subsection{Limitations}
The employed object detection tool was constrained to a given set of known objects (see \ref{section:expe_choices}) and its limited computation speed narrowed this study to the grasping of immobile objects. Moreover, observation tools inconsistent efficiency (see \ref{section:recordings_pre_processing}) led to the exclusion of a significant amount of trials, which may raise concerns about the handling of failed detections in real-life implementation. 
Nevertheless, one may be confident that the computer vision research field, being actively driven by robotics, will soon provide tools enhanced in universality, speed and accuracy. In addition, fine tuning hand detection on a user would be relevant in terms of specificity and reliability. In any case, even if these issues are related to processes upstream of i-GRIP, adapted safeguards would need to be implemented to prevent any action if hands are lost for a too long period.

This experiment was meant as a proof of concept for i-GRIP and was, as such, restricted to able-bodied participants. 
While the simulated movement type (see \ref{section:expe_proc}) was meant to mimic impaired motion, no specific instructions were given regarding movements velocity. 
Fig. \ref{fig_eval_participants}(iii) shows that i-GRIP performed slightly worse on simulated grasping for target identification but equally for grip selection, giving hope in its applicability for impaired users. 
Moreover, it is most likely that impaired users would perform slower movements, suggesting that i-GRIP's margins would be greater than those reported in Table \ref{table_target_identification}, giving time to an assistive device to trigger its actions during motion. 
Yet, these very actions could possibly result in different kinematics than the ones analysed in this work \cite{moullet}, and further studies involving impaired participants using real assistive devices will be required.

Finally, this work didn't set any confidence threshold on target selection from which the device grasping action would be triggered. Setting this value, along with tuning other i-GRIP's parameters would allow for a personalized compromise between reactivity and reliability, according to a user's preferences, pathology and assistive device.

\subsection{Additional observations}
As previously stated, the used computer vision tools exhibited inconsistent performances. Although it was not the main focus of this study, in the absence of a straightforward argument about the homogeneity of these inconsistencies in function of camera placement, its influence on trial validation was evaluated. 
Video trial validation rates presented in Fig. \ref{fig_eval_participants}(A) show that a device placement contralateral to the moving hand seems to lead to slightly more efficient hand and object detections throughout trials. Nonetheless, these results suggest that the video trials selected for i-GRIP's evaluation were overall homogeneously representative of the trials diversity.

The reported overall success rates for target identification and grip selection show little to no influence of camera placement over i-GRIP performances. This suggests that i-GRIP may be fed with inputs from both hands extracted from a single recording device, which can be placed arbitrarily according to any other design criterion. 

\section{CONCLUSIONS}
This paper stands as a proof of concept for a collaborative and generic grasping movement intention estimator. Further studies will be needed to assess i-GRIP in real-life use cases with users with upper limb motor deficiencies and for diverse conditions and assistive devices. Although our goal was to keep i-GRIP as generic as possible, its many parameters could be fine-tuned to enhance and tailor its performances to specific situations and users' preferences.

\addtolength{\textheight}{-12cm}   

\end{document}